\documentclass[nofootinbib,a4paper,fleqn,aps,prd,10pt,superscriptaddress,reprint,showkeys,showpacs]{revtex4-1}

\usepackage{times}
\usepackage{amsfonts}
\usepackage{amsmath}
\usepackage{amssymb}
\usepackage{natbib}
\usepackage{graphicx}
\usepackage{enumitem}
\usepackage{xcolor}
\usepackage[colorlinks=true,linkcolor=blue]{hyperref}
\usepackage[outdir=./]{epstopdf}
\usepackage[normalem]{ulem}
\usepackage{longtable}
\usepackage{supertabular}

\def\araa{ARAA}
\def\aap{A\&A}

\def\apj{ApJ}
\def\apjl{ApJL}
\def\apjs{ApJS}
\def\apss{Astrophysics and Space Science}
\def\jcap{JCAP}
\def\mnras{MNRAS}
\def\na{New Astronomy}

\begin{document}

\title{{On the time dependency of $a_0$
}}

\author{Antonino Del Popolo}
\email{antonino.delpopolo@unict.it}
\affiliation{Dipartimento di Fisica e Astronomia, University of Catania, Viale Andrea Doria 6, 95125 Catania, Italy}
\affiliation{Institute of Astronomy, Russian Academy of Sciences, Pyatnitskaya str. 48, 119017 Moscow, Russia}
\affiliation{Institute of Astronomy and National Astronomical Observatory, Bulgarian Academy of Sciences, 72 Tsarigradsko Shosse Blvd., 1784 Sofia, Bulgaria}

\author{Man Ho Chan}
\affiliation{Department of Science and Environmental Studies, The Education University of Hong
Kong, Tai Po, New Territories, Hong Kong, China}
%\affiliation{PDAT Laboratory, Department of Physics, K.N. Toosi University of Technology, P.O. Box 15875-4416, Tehran, Iran}

%
%\author{Author~\surname{a}}%
%\affiliation{
%%Dipartimento di Fisica e Astronomia, University of Catania, Viale Andrea Doria 6, 95125, Catania, Italy
%}
%\affiliation{
%%Institute of Astronomy, Russian Academy of Sciences, Pyatnitskaya str. 48, 119017
%%Moscow, Russia
%%INFN sezione di Catania, Via S. Sofia 64, I-95123 Catania, Italy
%}
%
%\email[Corresponding author: ]{adelpopolo@oact.inaf.it}

%
%\author{Author~\surname{b}}%
%\affiliation{
%Jodrell Bank Centre for Astrophysics, School of Physics and Astronomy, The University of %Manchester, Manchester, 
%M13 9PL, U.K.
%}
%\email[]{francesco.pace@manchester.ac.uk}
%
%\author{David~F.~\surname{Mota}}%
%\affiliation{%
%Institute of Theoretical Astrophysics, University of Oslo, P.O. Box 1029 Blindern, N-0315 Oslo, %Norway 
%}
%\email[]{d.f.mota@astro.uio.no}
%

\label{firstpage}

\date{\today}

\begin{abstract}
In this paper, we test one of the predictions of the Scale Invariant Vacuum (SIV) theory on MOND. According to that theory, 
MOND's acceleration $a_0$ is not a universal quantity but depends on several parameters, including time. To check this prediction, we compare the dependency of $a_0$ from redshift, using the values of $a_0$ obtained in
\citet{Marra2020} by carrying a Bayesian inference for 153 galaxies of the SPARC sample. Since this sample does not contain galaxies at large redshift, we have estimated $a_0$ from the data in \citet{Nestor}. The SPARC sample, for small values of the redshift, in the redshift range $0.00032-0.032$ gives a correlation with $z$, while the \citet{Nestor} data, in the higher redshift range $0.5-2.5$ gives an anti-correlation with the redshift $z$. {Both samples show a dependency of $a_0$ from $z$, although the uncertainties involved are large, especially for the high-redshift galaxies. The combined sample gives an overall correlation of $a_0$ with $z$.} The different behavior at low and high redshift can be related to a change of the $a_0(z)$ with redshift, or to the lower precision with which the high-redshift value of $a_0$ are known. 
\end{abstract}



\maketitle

\section{Introduction}

In the past decades, the $\Lambda$CDM model showed to be a very good model in predicting 
observations on cosmological scales \cite{2011ApJS..192...18K,DelPopolo2007,2013AIPC15482D}, except some drawbacks, and on intermediate scales (\citep{Spergel,Kowalski,Percival,Komatsu,2013AIPC15482D,2014IJMPD2330005D}. 
Some of the drawbacks on cosmological scales are the cosmic coincidence problem \citep{Velten2014}, the cosmological constant problem \citep{Weinberg, Astashenok}, the Hubble tension, namely a discrepancy in $H_0$, the current value of the Hubble parameter, in which the value fitted using the CMB is different from that in the local universe using supernovae and stars
\citep{DiValentino2021}. The CMB also shows an unexpected cold spot \citep{cruz,cruz1,cruz2}, a quadrupole-octupole alignment \citep{schwa,copi,copi1,copi2,copi3}, a hemispherical asymmetry \citep{eriksen,hansen,jaf,hot,planck1,akrami}, and another tension concerning the growth rate of perturbations, $\sigma_8$, when using different measurements \citep{maca,raveri}. 
Another problem of the model is that the constituents of dark matter have never been detected, and there are no clues on what 
``dark energy" could be.
In the scale $1-10$ kpcs, the $\Lambda$CDM model suffers from the so called ``small scale problems, like 
the ``Cusp/Core" problem 
\citep{Moore1994,Flores1994,Burkert1995,deBloketal2003,Swatersetal2003,DelPopolo2009}\citep{DelPopoloKroupa2009,2012MNRAS419971D,DelPopoloHiotelis2014,nfw1996,nfw1997,Navarro2010} 
the ``missing satellite problem" \citep{Klypin1999,Moore1999}, etc. 
Apart drastic solutions to solve the ``small scale problems", like modifying the particles constituting dark matter (\citep{2000ApJ542622C,2001ApJ551608S,2000NewA5103G,2000ApJ534L127P}), or modifying the power spectrum (e.g. \citep{2003ApJ59849Z}), astrophysical solutions have been proposed based on the role of baryons (\citep{1996MNRAS283L72N,Gelato1999,Read2005,Mashchenko2006,Governato2010,El-Zant2001,El-Zant2004,2008ApJ685L105R,DelPopolo2009,Cole2011,2012MNRAS419971D,Saburova2014,DelPopoloHiotelis2014}).
Another solution proposed to solve some of the drawbacks of the $\Lambda$CDM model is modifying 
the theory of gravity (\citep{1970MNRAS1501B,1980PhLB9199S,1983ApJ270365M,1983ApJ270371M,Ferraro2012}). In some cases 
these theories do not even need dark matter and dark energy, like for example the MOND theory \citep{Milgrom1986}.
Another theory without the need of the two quoted components is the Scale Invariant Vacuum (SIV) theory, based on the hypothesis that the macroscopic space is scale invariant. 

In \citep{Maeder2023}, the relations between SIV and MOND were studied. The author found the deep-MOND limit equations are reobtained, and that contrarily to MOND assumption, $a_0$ is not a universal constant but is time dependent.
In the present paper, we want to check the time dependency prediction of $a_0$. For this aim, we will use the SPARC sample 
and \citet{Nestor} data, to calculate the values of $a_0$ and their relation with the redshift $z$. 
The paper is organized as follows. Section II describes SIV, and its connection with MOND. In Section III, we discuss the data and statistical analysis. Section IV and V are devoted to discussion and conclusion, respectively.

\section{$a_0$ time dependence}
%\section{SIV's MOND approximations}

{One of the central point in MOND is the existence of an universal acceleration, namely $a_0$, that distinguishes between the Newtonian and non-Newtonian behavior of a system. Nevertheless MOND theory claims that $a_0$ is universal and constant with a value $a_0=1.2\times 10^{-8}$ cm/s$^2$. Several studies have shown that this is not the case \citep{Saburova2014,Rodrigues2018a,Rodrigues2018b,Rodrigues2020,Marra2020,Zhou2020,DelPopolo2023a}. However, none of the previous studies has considered the possibility that $a_0$ is time-dependent. Based on the claims in \citep{Maeder2020,Maeder2023} that $a_0$ should be a time-dependent quantity, and the relation between SIV and MOND theories (see the Appendix), in this paper, we would like to check this possibility.  }

\begin{figure*}[!ht]
 \centering
 \includegraphics[width=10cm,angle=0]{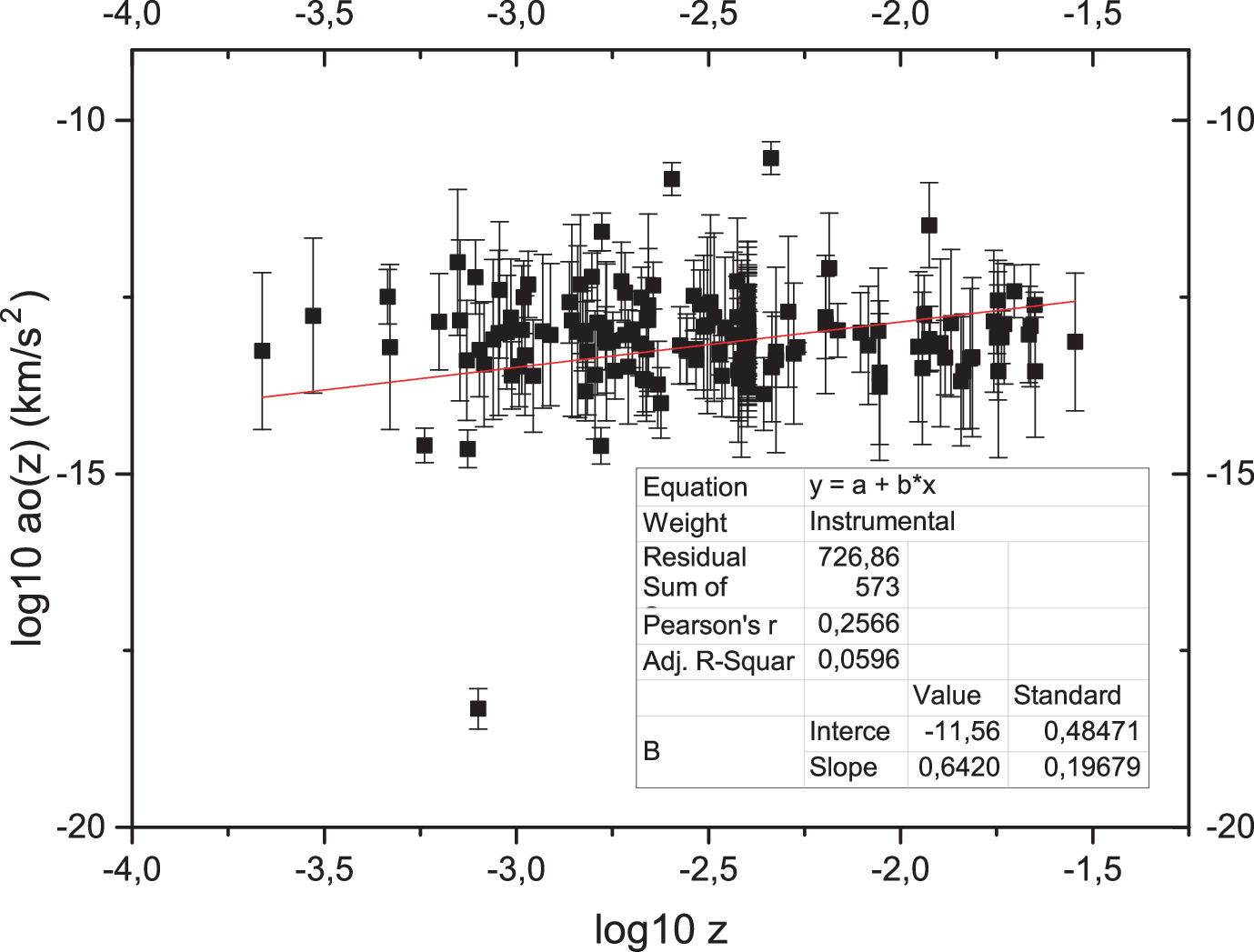}
\includegraphics[width=10cm,angle=0]{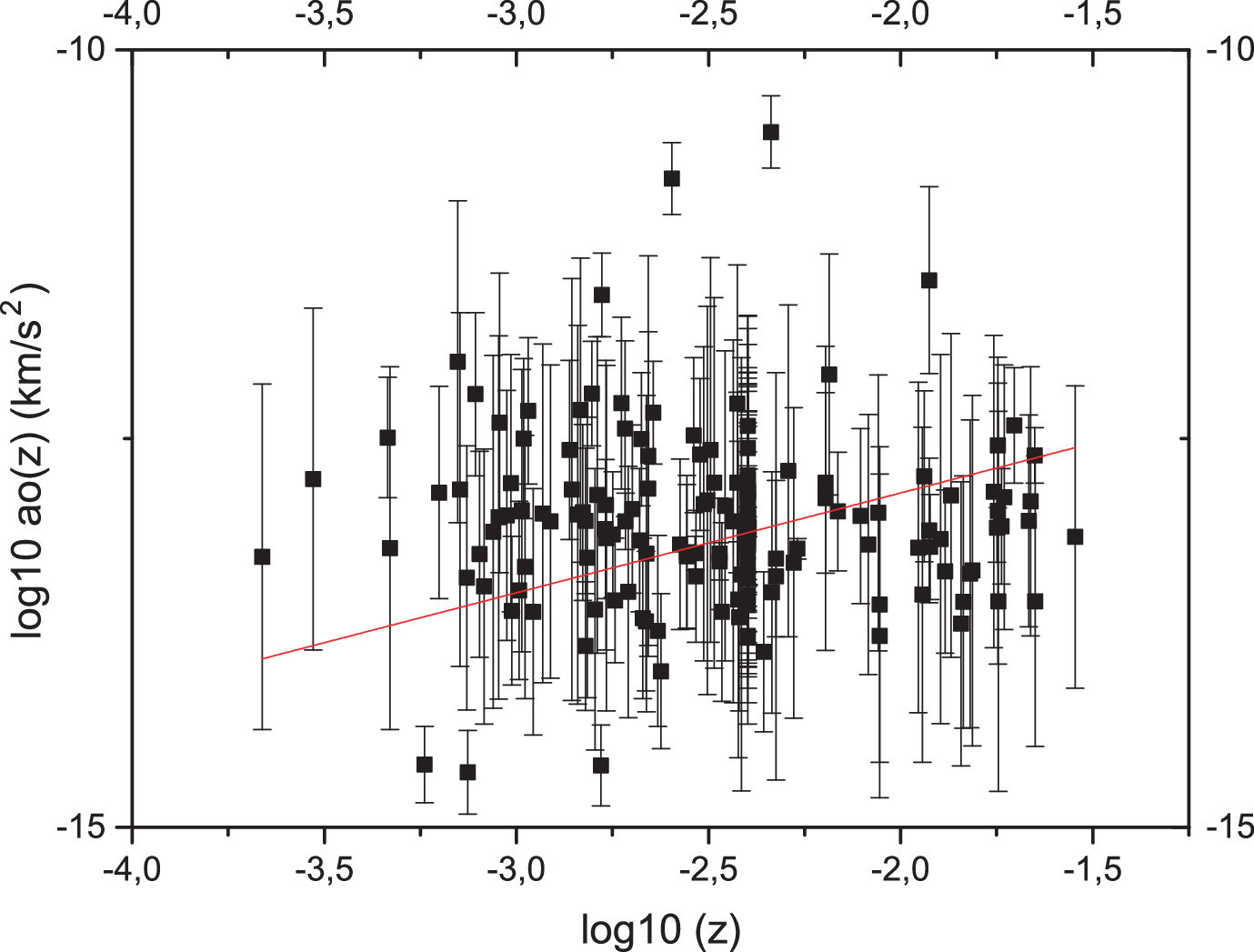}
 \caption[justified]{
Upper panel: The black dots with error bars represent the values of $a_0$ (with $1\sigma$ standard deviation) against redshift 
 $z$ obtained from the SPARC data. The red solid line indicates the non-linear fit to the data points. Bottom panel: same as the top panel but with a different ordinate axis scale.
 %  for the small-redshift group (the SPARC sample), medium-redshift group ($z=0.5-1.0$), and the high-redshift group 
%($z=1.5-2.5$) of galaxies. 
}
 \label{fig:comparison}
\end{figure*}

%
%\begin{figure*}[!ht]
% \centering
%\includegraphics[width=15cm,angle=0]{Origin_MOND_err_NonLog.eps}
%\includegraphics[width=15cm,angle=0]{Origin_MOND_err_NonLog1.eps}
% \caption[justified]{...
%}
% \label{fig:comparison}
%\end{figure*}
%

\section{Data and statistical analysis}

As mentioned above, in the paper, we use the SPARC ($\it Spitzer$ Photometry and Accurate Rotation Curves) dataset \citep{Lelli2016} for analysis. The dataset is constituted by 175 late-type galaxies with surface photometry at 3.6 $\mu \rm m$, and with high quality rotation curves obtained from $\rm HI/H\alpha$ studies. 
The gas mass is provided by 21 cm observations. The morphologies present in SPARC span from S0 to Irr. 
Almost all SPARC galaxies have a disc structure, with some having also bulges. The baryon component is constituted by the disc, bulge, and gas components. 
{The SPARC dataset, with some quality cuts, was used by \citep{McGaugh2016} to derive the radial acceleration relation (RAR), and by \citep{Marra2020} to perform a test on MOND, based on \citep{Rodrigues2018a,Rodrigues2018b,Rodrigues2020}. After the quoted cuts described in \citep{McGaugh2016,Marra2020}, 153 galaxies are left.  
%
%The Newtonian acceleration, since dark matter is not considered, has  the baryonic component. This is subdivided into stellar %and atomic gas components. The centripetal Newtonian acceleration, coming from the baryonic distribution, can be written as 
%\begin{equation} \label{newtdecompA}
%a_N 
%= \Upsilon_b a_b + \Upsilon_d a_d+ a_{\rm gas} \,.
%\end{equation}
%In the previous equation, $a_b$ and $a_d$ represents the bulge and disk contributions, respectively, to the centripetal %acceleration for mass-to-light ratios ($\Upsilon_b$ and $\Upsilon_d$) equal to one. Equivalently, we have the velocity %decomposition as, 
%\begin{equation} \label{newtdecompV}
%V_N^2 
%= \Upsilon_b|V_b| V_b + \Upsilon_d|V_d| V_d + |V_{\rm gas}| V_{\rm gas} \,.
%\end{equation}
%
\citep{Marra2020} carried a Bayesian inference for the quoted 153 galaxies. 
They adopted for each galaxy the Gaussian likelihood
\begin{align}
\mathcal{L}(\theta ) =
|2 \pi \Sigma|^{-1/2}
e^{-\chi^{2}(\theta)/2} \,,
\end{align}
where $\Sigma$ is the covariance matrix described in \citep{Marra2020}, and $\chi^2 = \chi^2(A_0,Y_b,Y_d, D, I )$, with 
\begin{equation} \label{chi2fun}
\chi^{2}= \sum_{i=1}^{N} \left ( \frac{ V_M(R_i,A_0, Y_b, Y_d, D) -V_{C, i} \frac{\sin I_0}{\sin I} }
{\sigma_{V, i} 
\frac{\sin I_0}{\sin I} }\right)^2 \, .
\end{equation}
In Eq. (\ref{chi2fun}), $N$ is the number of data points relative to the given galaxy data, $V_M$ is the circular velocity of the model, $R_{i}$ represents the galaxy radius at which the circular velocity $V_{C,i}$ was measured, $\sigma_{V,i}$ is the corresponding error, and $I_0$ is the inclination, given together with $V_{C, i}$ and $\sigma_{V, i}$ by SPARC.
The model velocity $V_M$ is obtained from MOND using the interpolating function
\begin{equation} \label{int}
{\bf a}= \frac{{\bf a_N}}{1-e^{-\sqrt{a_n/a_0}}}
%\boldsymbol{a} = \frac{\boldsymbol{a_{\mscript{N}}}}{1 - e^{-\sqrt{a_\mtiny{N}/a_0}}} \,. 
\end{equation}
the Newtonian velocity, $V_N$,  
\begin{equation} \label{newtdecompV}
V_N^2 = \Upsilon_b|V_b| V_b + \Upsilon_d|V_d| V_d + |V_{\rm gas}| V_{\rm gas} \,.
\end{equation}
Here, the subscripts $b$ and $d$ stand for bulge and disc respectively, and $\Upsilon_b$ and $\Upsilon_d$ are the corresponding mass-to-light ratios, with the distance correction given in Eq. 5 of \citep{Marra2020}. The priors chosen, the quality cuts, and the determination of the global best value are described in Sections 4.2, 4.3, and 4.4 of \citep{Marra2020} respectively.}

Moreover, based on the data of the high-redshift galaxies in \citet{Nestor}, we can get a sample of high-redshift galaxies for analysis. These samples can represent the average values of $a_0$ at high redshift ($z=0.5-2.5$). However, it is very difficult to obtain the value of $a_0$ for each of these high-redshift galaxies because the data of the galactic rotation curve profiles are too few. Fortunately, there are some galaxies in which the gravitational acceleration at the effective radius $R_e$ is smaller than the characteristic acceleration scale $a_0$. That means the position at $R_e$ for these galaxies are approximately located at the deep-MOND regime. In this regime, we have \citep{Sanders}
\begin{equation}
V_c^4=GM_Ba_0,
\end{equation}
where $V_c$ is the rotation velocity at $R_e$ and $M_B$ is the total baryonic mass within $R_e$. From Table 3 of \citet{Nestor}, we find that there are 17 galaxies in which the gravitational acceleration is smaller than the typical $a_0=1.2\times 10^{-8}$ cm/s$^2$ at $R_e$. We plot the values of $a_0$ as a function of $z$ in Fig.~2. We can see a very small decreasing trend exists in the plot when $z$ is larger. 
%Simple regression gives $\log a_0=-0.025z-12.6$, where $a_0$ is in the unit of km/s$^2$. 
Unfortunately, the uncertainties of the high-redshift rotation curve data are very large. Therefore, the result here can be used as a reference only.

%
%To get a better comparison, we group the SPARC sample as a low-redshift group ($z<0.1$), the galaxies with $z=0.5-1.0$ as a %medium-redshift group, and the galaxies with $z=1.5-2.5$ as a high-redshift group. By taking the average value of $a_0$ in %each group, we plot the values of $a_0$ as a function of $z$ in Fig.~4. A clear decreasing trend of $a_0$ when $z$ is larger %can be seen. 
%

\begin{figure*}
\vskip 10mm
 \includegraphics[width=10cm,angle=0]{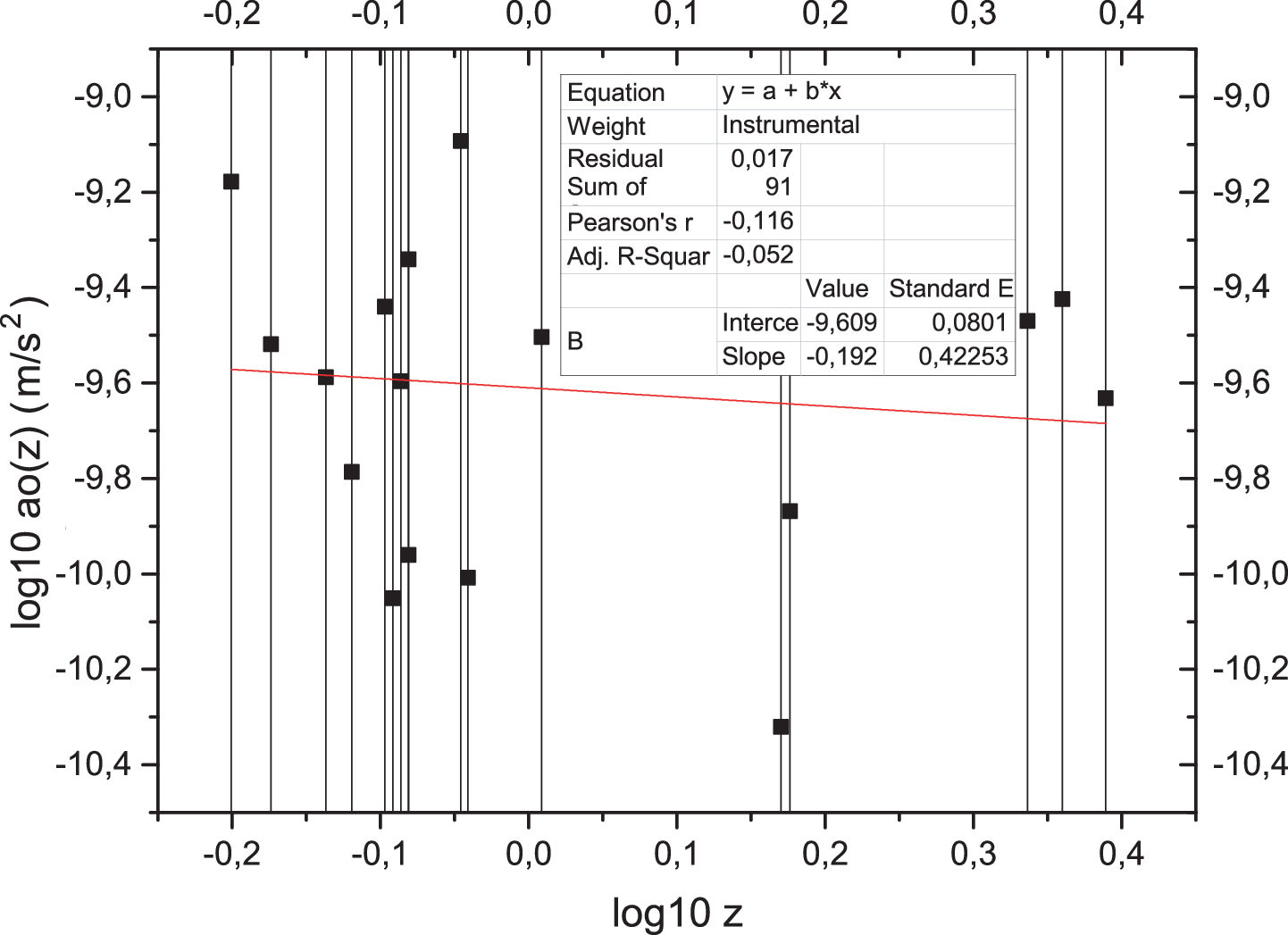}
 \includegraphics[width=10cm,angle=0]{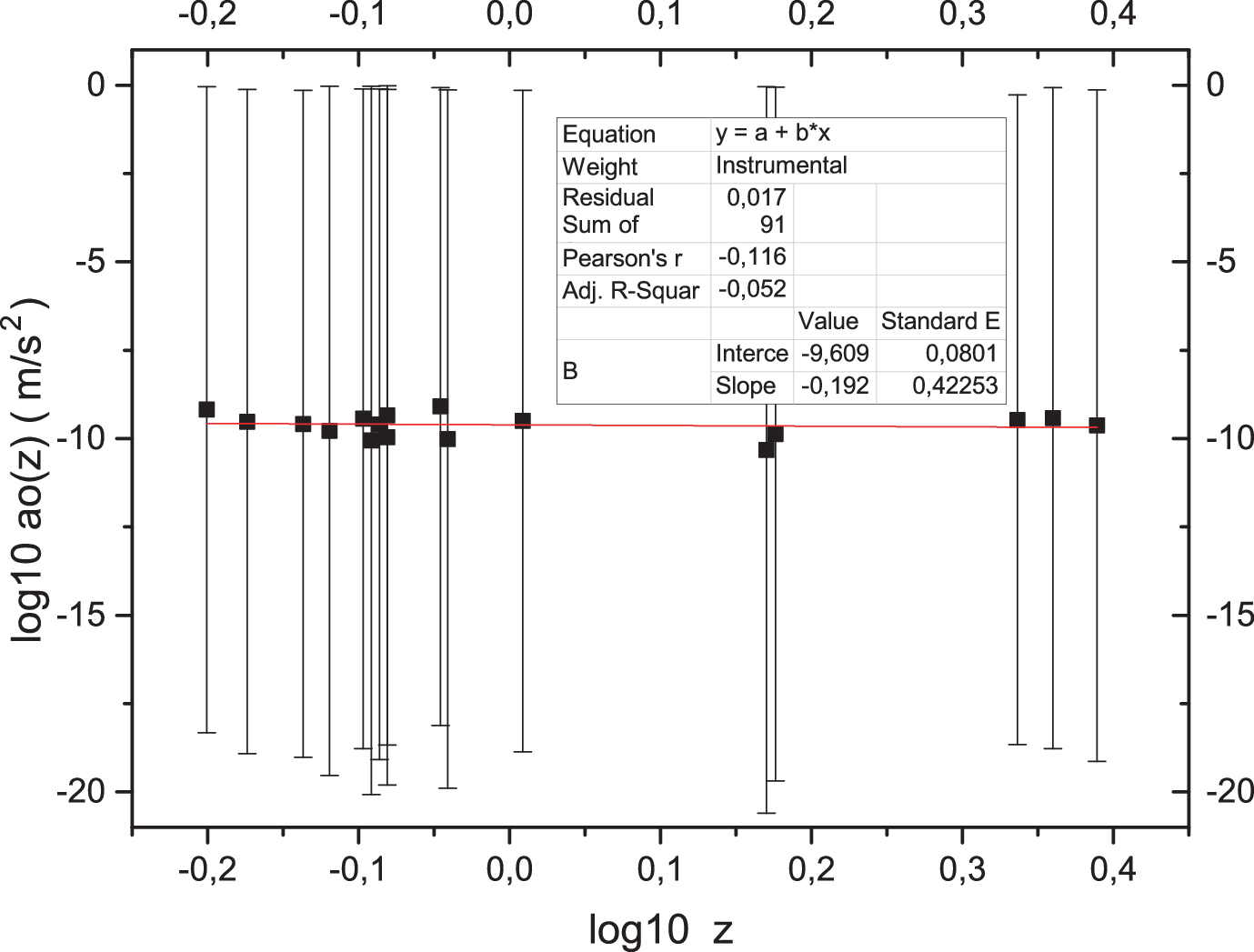}
 \caption{Top panel: The black dots with error bars represent the values of $a_0$ (with $1\sigma$ standard deviation) against redshift  $z$ obtained from the  \citet{Nestor} data. The red solid line indicates the non-linear fit to the data points. Bottom panel: same as the top panel but with a different ordinate axis scale.
 %  for the small-redshift group (the SPARC sample), medium-redshift group ($z=0.5-1.0$), and the high-redshift group 
%($z=1.5-2.5$) of galaxies.
}
\vskip 10mm
\end{figure*}

%
%\begin{figure}
%\vskip 10mm
% \includegraphics[width=80mm]{trend.eps}
% \caption{The black dots with error bars represent the values of $a_0$ (with $1\sigma$ standard deviation) against redshift 
%$z$ for the small-redshift group (the SPARC sample), medium-redshift group ($z=0.5-1.0$), and the high-redshift group 
%($z=1.5-2.5$) of galaxies.}
%\vskip 10mm
%\end{figure}
%

%\bibitem[Nestor et al. (2023)]{Nestor} Nestor Shachar, A. {\it et al.}, 2023, Astrophys. J. 944, 78.
%\bibitem[Sanders \& McGaugh (2002)]{Sanders} Sanders, R. H. \& McGaugh, S. S., 2002, Annu. Rev. Astron. Astrophys. 40, %263-317.

\begin{figure*}
\vskip 10mm
 \includegraphics[width=10cm,angle=0]{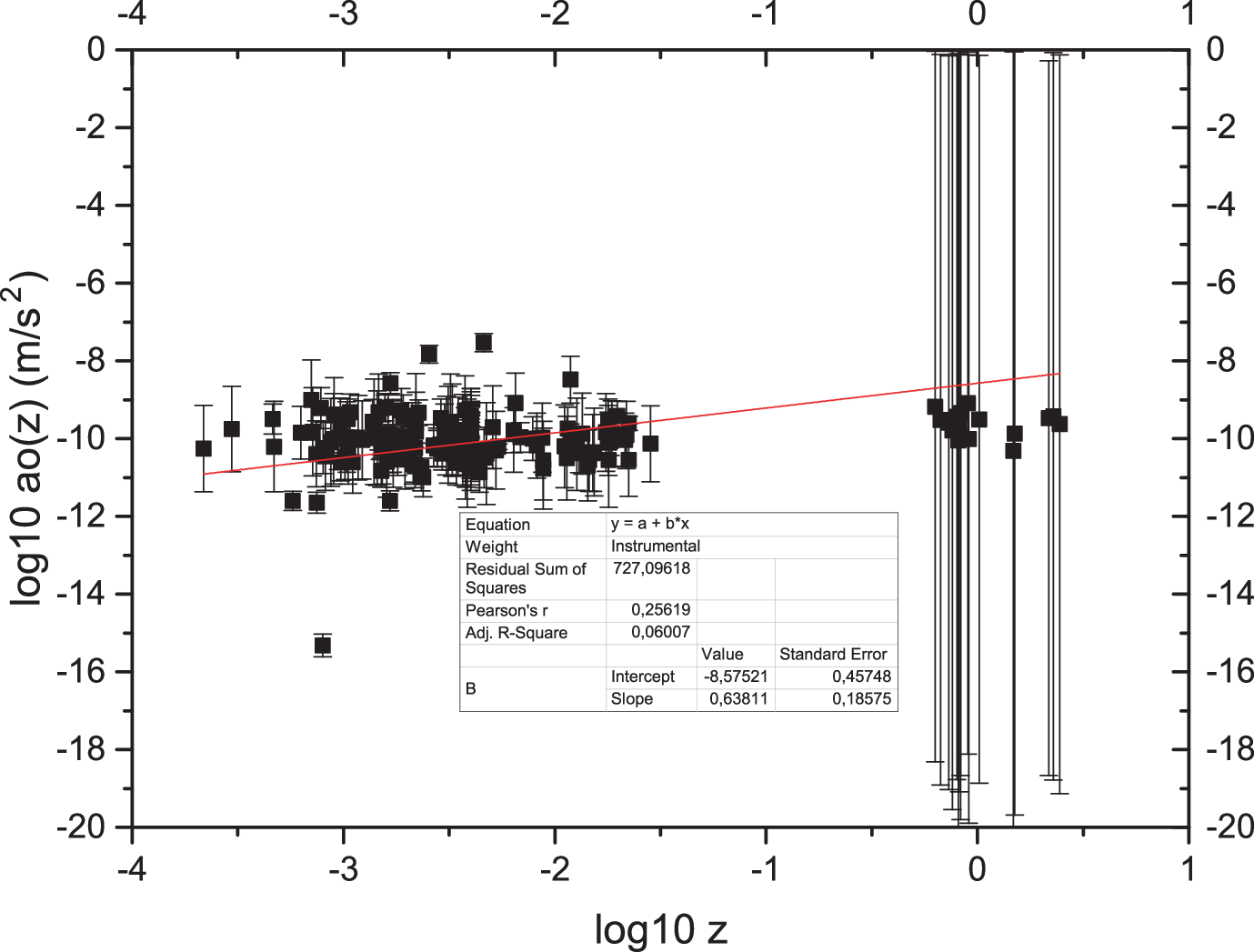}
 \caption{The black dots with error bars represent the values of $a_0$ (with $1\sigma$ standard deviation) against redshift 
 $z$ obtained from all the data we have. The red solid line indicates the non-linear fit to the data points.
 %  for the small-redshift group (the SPARC sample), medium-redshift group ($z=0.5-1.0$), and the high-redshift group 
%($z=1.5-2.5$) of galaxies.
}
\vskip 10mm
\end{figure*}

\section{Discussion}  

The MOND paradigm is successful in describing rotation curves, and other astrophysical aspects. It is based on the idea that Newtonian mechanics is valid only for higher values of gravitational acceleration, and for lower values it must be modified. 
The threshold between the regimes in which Newtonian mechanics, and MOND are to be applied is related to an universal parameter of the model, the acceleration $a_0$. Several studies have shown that in reality $a_0$ is not a universal constant \citep{Saburova2014,Rodrigues2018a,Rodrigues2018b,Rodrigues2020,Zhou2020,DelPopolo2023a} depending on several parameters. 
Prompted by the paper of \citep{Maeder2023}, showing that the SIV theory reduces to MOND prediction of the acceleration, in certain limits, and that the acceleration $a_0$ of MOND cannot be a universal constant but a parameter that changes with time, we checked this possibility. Using the values of $a_0$, and their errors obtained by \citep{Marra2020}, which are an improvement on that obtained in \citep{Rodrigues2018a} using Bayesian statistics, and the redshift of the same galaxies, we studied the case of $a_0$ being a function of redshift. Then, we performed a non-linear fitting using the Levenberg-Marquardt algorithm, on the \citep{Marra2020} data, in order to account for errors in the dependent variable. The result is shown in Fig. 1. the top and bottom panel represent the same fit considering different scale for the box. The plot clearly shows a correlation between $a_0$, and redshift. The Pearson's r coefficient is 0.25661. The other estimated parameters of the fit are written directly on the Fig. 1. It is interesting to note that the redshift range is 0.00032-0.032, namely even for low redshift the correlation is present. In order to see if the correlation continues to persist going towards larger redshifts, we used the the data of the high-redshift galaxies in \citet{Nestor}. These samples have much larger redshift than SPARC, with redshift in the range $z=0.5-2.5$. While in the case of the SPARC sample the values of $a_0$ were obtained from the rotation curves using solid statistical methods, in the case of \citet{Nestor} sample, we do not have the rotation curves of the galaxies, and we cannot obtain $a_0$ with the methods applied to the SPARC sample. Since the gravitational acceleration at the effective radius $R_e$, for some galaxies, is smaller than $a_0$, and they are located in the deep-MOND regime, and in this case we can get an approximate value of $a_0$ from the relation $V_c^4=GM_Ba_0$, where $M_B$ is the total baryonic mass within $R_e$, and $V_c$ is the rotation velocity at $R_e$. Clearly this is just an estimate that cannot be precise as the $a_0$ obtained from SPARC with Bayesian statistics. We used 17 galaxies from \citet{Nestor}, those having 
the gravitational acceleration smaller than the typical $a_0=1.2\times 10^{-8}$ cm/s$^2$ at $R_e$. Then, we fitted with the same method used for the SPARC galaxies (Fig. 1), the $a_0$s estimated from \citet{Nestor} data. The result is shown in Fig. 2. The fit shows a anti-correlation between $a_0$ and the redshift weaker than in the case of the SPARC sample. The Pearson's r coefficient is in this case -0.116, and the other parameters of the fit are written on Fig. 2. As reported, Fig. 2 shows an anti-correlation of $a_0$ with $z$, while Fig. 1 a correlation between the same quantities. The reason leading to this could be several. As Fig. 2 shows, the error bars in $a_0$ are very large, and the fitting algorithm is sensitive to the errors. Small changes in the errors could produce a different behavior on the fit. 

{Note that the values of $a_0$ obtained in Fig.~1 and in Fig.~2 are following different methods. Generally speaking, using Bayesian method to get the values of $a_0$ would be more reliable. However, for the high-redshift galaxies, the data points of the rotation curves are too few. The number of variables involved in the Bayesian analysis would be larger than or almost the same as the number of data points. In this situation, the results from the Bayesian analysis would not be reliable. Therefore, we need to use another method (i.e. the deep-MOND regime approximation) to get the values of $a_0$. Although the methods used in the two samples are different, the values of $a_0$ found in the two different samples originate from the same MOND framework. Nevertheless, the values of $a_0$ obtained for the high-redshift galaxies are not known with the precision of those of SPARC, and probably the result in Fig. 2 is less reliable than that in Fig. 1.} A last possibility is that the $a_0$ dependence on redshift is non-linear: it increases reaches a maximum and then decreases. In order to evaluate these possibilities, it would be necessary to find other galaxies at high redshift for which the rotation curves are known. 
{In any case, we are more interested in the overall trend of $a_0$ with redshift. For this reason, we plotted all the data on the same plot in Fig. 3, and fitted again the data. The final result is that there is a correlation between $a_0$, and redshift. The Pearson's r coefficient is 0.26.}
Apart from the results using galaxies, some previous studies using the data of galaxy clusters (HIFLUGCS and CLASH samples) have obtained larger values of $a_0$ by analysing the radial acceleration relation \citep{Chan2020,Tian2020}. The redshift ranges of the data samples HIFLUGCS and CLASH are $z\sim 0.01-0.2$ and $z \sim 0.2-0.7$ respectively, which are in between the SPARC sample and the high-redshift galaxies used in this study. The values of $a_0$ obtained are $a_0 \sim 10^{-7}$ cm/s$^2$, which are larger than that shown in galaxies. Therefore, these show that the value of $a_0$ might not be a constant. We anticipate that more data ranging from galaxies to galaxy clusters with different redshifts can further verify our proposal.

\section{Conclusion}

According to the $\Lambda$CDM model, universe is dominated by dark matter, and dark energy. {Nevertheless, this dark matter component remains undetected from many different experiments.} Concerning dark energy, the situation is even worse since we do not really know what dark energy is composed of. Moreover, the $\Lambda$CDM model shows some drawbacks especially at small scales. In order to solve those drawbacks in the $\Lambda$CDM model, several authors have proposed to modify the theory of gravity (\citep{1970MNRAS1501B,1980PhLB9199S,1983ApJ270365M,1983ApJ270371M,Ferraro2012}). In some of these modified gravity theories, dark matter and dark energy are not needed. One of these theories is 
the Scale Invariant Vacuum (SIV) theory, based on the hypothesis that the macroscopic space is scale invariant. One of the predictions of this theory is that the quantity $a_0$ in MOND is time-dependent. In order to check this prediction, 
we compared the dependency of $a_0$ from redshift, using two groups of data. The first one, of higher quality, and low values of $z$, is obtained in \citet{Marra2020} by carrying a Bayesian inference for 153 galaxies of the SPARC sample. The second one, of lower quality but larger redshift is obtained from the data in \citet{Nestor}.
{Both sample shows a dependency $a_0$ from redshift, although the involved uncertainties are large, especially for the high-redshift galaxies.} The $a_0$ obtained from the SPARC sample shows a correlation with $z$, at small redshifts, while the $a_0$ obtained from the data of \citet{Nestor} shows an anti-correlation with redshift, at higher $z$. In other terms, both samples show a dependency of $a_0$ from redshift. The difference of behavior at small and larger redshift can be due to a change of the behavior of $a_0$ with $z$, or to the lower quality of the high redshift data. The combined data give a correlation of $a_0$ with $z$.

\newpage
{\bf APPENDIX}

Scale invariance was considered by Weyl, and Eddington \citep{Eddington1923} in order to account for gravitation, and electromagnetism by the space-time geometry. It was initially abandoned because particle properties were dependent on its past worldline. Considering the Weyl Integrable Geometry (WIG), the problem is solved \citep{Dirac1973,Canuto1977}. By scale invariance one means that through a transformation of the line element like
\begin{equation}
ds^{'}=\lambda(x^\mu) ds,
\end{equation} 
the basic equations do not change. $\lambda(x^\mu)$ is the scale factor, $ds^{'}$ refers to general relativity (GR), while $ds$ refers to the WIG space. \citep{Bouvier1978} studied the WIG geometrical properties, and geodesics, while \citep{Maeder1979}
studied the weak field limit. Differently from GR is present a small additional acceleration in the direction of motion. Several studies, with positive results, have been performed on the rotation of galaxies, the dynamics of clusters of galaxies, on the growth of density fluctuation in the early universe, on inflation, on lunar recession (see \citep{Maeder2020,Maeder2023} for the references).

In \citep{Maeder2020,Maeder2023}, the SIV basic theoretical context, and the cosmological solutions are given. We are interested in the weak field approximation, and the MOND approximation. The weak field approximation was obtained by \citep{Maeder1979}, and can be expressed, in spherical coordinates, as
\begin{equation}
\frac{d^2 {\bf r}}{dt^2}=-\frac{G_t M(t)}{r^2} \frac{{\bf r}}{r}+\kappa(t) \frac{d {\bf r}}{dt}.
\end{equation}
The extra term is called {\it the dynamical gravity}. This term, which is proportional to the velocity, favors outwards motion during the expansion, and collapse during a contraction. In the case of a dust Universe, the conservation law imposes
$\varrho a^3 \lambda=const.$. This means that the inertial mass of a particle is not a constant, and moreover it depends on the factor $\lambda$. The time dependency of mass can be expressed as $M(t)  =   M(t_0)  (t/t_0)$. 
In the case $\Omega_{\mathrm{m}} =0.3$, the mass  at the Big-Bang was, $M(t_{\mathrm{in}}) =  \Omega_{\mathrm{m}}^{1/3} \, M(t_0)= 0.6694 \; M(t_0)$, where is $t_0=1$ at  present and  $t_{\mathrm{in}} = \Omega^{1/3}_{\mathrm{m}}$ at the origin. 
The timescale $\tau$ in years is defined as $\tau_0=13.8$ Gyr, and $\tau_{\rm in}=0$ at the Big-Bang. 
$\tau$, and $t$ are related by
\begin{equation}
\frac{\tau - \tau_{\mathrm{in}}}{\tau_0 - \tau_{\mathrm{in}}} = \frac{t - t_{\mathrm{in}}}{t_0 - t_{\mathrm{in}}}\, ,
\end{equation}
and
\begin{equation}
\tau \,= \, \tau_0 \, \frac{t- \Omega^{1/3}_{\mathrm{m}}}{1- \Omega^{1/3}_{\mathrm{m}}} \,  \quad \mathrm{and} \; \;
  t \,= \, \Omega^{1/3}_{\mathrm{m}} + \frac{\tau}{\tau_0} (1- \Omega^{1/3}_{\mathrm{m}}) \,,
\label{tau}
\end{equation}
At present time, $\tau_0$, the modified Newton's equation is given by
\begin{equation}
 \frac {d^2 \bf{r}}{d \tau^2}  \, = \, - \frac{G \, M(\tau_0)}{r^2} \, \frac{\bf{r}}{r}   + \frac{\psi_0}{\tau_0}   \frac{d\bf{r}}{d\tau} \, \,.  
\label{newton}
\end{equation}
where
\begin{equation}
G=G_t (\frac{dt}{d \tau})^2
\end{equation}
and 
\begin{equation}
\psi(\tau) \, = \, 
%\frac{ 1}{\left[\frac{\Omega^{1/3}_{\mathrm{m}}}{1-\Omega^{1/3}_{\mathrm{m}}}+1 \right]} \, =
\frac{t_0-t_\mathrm{in}}{ t_\mathrm{in} +   \frac{\tau}{\tau_0} (t_0-t_\mathrm{in}) }\, ; \; \mathrm{and} \;
\psi_0=\psi(\tau_0) \, =1-\Omega^{1/3}_{\mathrm{m}} \,.
\label{psi}
\end{equation} 

In the case of galaxies, having rotation periods of some hundred years, $\lambda$, and $M$ can be considered constant at the level of 1\%. In Eq. (\ref{newton}) the dynamical gravity, namely the term proportional to $\kappa$ disappears. Considering the transformations $r = \lambda \; r' $  and  $t = \lambda \; t'$, and applying to the Newton equation  
%with a constant $\lambda$ (and thus $M$)
expressed  in the prime coordinates gives,
   \begin{equation}
     \frac{d^2r'}{dt'^2} \, = \, - \frac{GM}{r'{^2}} \,  \equiv  \, g'_N.
     \end{equation}
In the system $(r, t)$ the total acceleration $g$ is,
 \begin{equation}
 g\, = \, \frac{d^2r}{dt^2} \,=\,\frac{1}{\lambda} \, \frac{d^2 r'}{dt'^{2}}\,. 
 \end{equation}
The accelerations $g_N$ and $g'_N$ are related by,
 \begin{equation}
g_N \, \equiv \, - \frac{G \, M}{r^2} \, = \,- \frac{1}{\lambda^2}  \frac{G \, M}{r'^2}\, \equiv   \frac{1}{\lambda^2} \,g'_N\, .
 \end{equation}
Then the total acceleration is
 \begin{equation}
 g \, = \,\frac{d^2r}{dt^2} \,=\,\frac{1}{\lambda} \, \frac{d^2r'}{dt'^2}\, = \, \frac{1}{\lambda} \, g'_N \, = \, {\lambda} \, g_N \, ,
 \end{equation}
and finally
 \begin{equation}
g \, = \,  \frac{d^2r}{dt^2} \,= \, {\lambda} \, g_N \,  = \, \left(\frac{ g'_N}{g_N}\right)^{1/2}g_N \, = \left(g'_N \, g_N \right) ^{1/2} \, .
 \end{equation}
In the deep-MOND regime limit we have 
\begin{equation}
g=\sqrt{a_0 g_N}
\label{deep}
\end{equation} 
and then comparing, $g'_N=a_0$. So, in the initial approximation of constant $\lambda$, and $M$, the scale invariant theory leads us to the deep-MOND limit, and moreover $a_0$ is time dependent like $\lambda$ (in the general case). 
Another way to obtain the MOND behavior from Eq. (\ref{newton}) is to consider low gravities. In this case, one can show that
\begin{equation}
g \, = g_{\mathrm{N}}\left[1+  \frac{\sqrt{2}\,  \psi_0}{\xi} \,
\left(\frac{g_{\mathrm{c}}}{g_{\mathrm{N}}}\right)^{1/2} \right]\,.
\end{equation}
where $g_c$ is the mean gravity at the edge of a sphere having the critical density, $\xi=\tau_0 H_0$, and $\psi_0$ depends on $\Omega_m$ (see Eq. (\ref{psi})).
For regions at large distances from the galactic center the previous equation reduces to
\begin{equation}
 g \, \simeq
 %\rightarrow  
 \,  \frac{\sqrt{2}\,  \psi_0}{\xi} \, \left( g_{\mathrm{c}} \, g_{\mathrm{N}} \right)^{1/2}\,
\label{g}
 \end{equation}
where
\begin{equation} \label{psi}
\frac{\sqrt{2}\,  \psi_0}{\xi}= \frac{\sqrt{2}\, (1- \Omega^{1/3}_{\mathrm{m}})}{H(\tau_0)\, \tau_0} =
  \frac{(1- \Omega_{\mathrm{m}})}{\sqrt{2}}\, ,
  \end{equation}
 So, in the weak field limit, Eq. (\ref{newton}) leads to Eq. (\ref{g}) which is equivalent to the deep-MOND limit, namely Eq. (\ref{deep}).  

In the case of SIV or $\Lambda$CDM models with $\Omega_m$ in the range $0.2-0.3$, we have

\begin{equation} 
 a_0  =  \frac{(1-\Omega_{\mathrm{m}})^2}{4} \, n\,  c \, H_0\, 
\end{equation}
or 
\begin{equation}
 a_0 =  \frac{n\, c \, (1-\Omega_{\mathrm{m}}) (1-\Omega^{1/3}_{\mathrm{m}})}{2\, \tau_0}
\end{equation}   
$n$ depends on the cosmology, and is defined in \citep{Maeder2023}. 
%
%It is defined as follows
%{\bf ??????
%A quantity $Y$, scalar, vector or tensor, which in a scale transformation changes  like $Y' \,= \, \lambda^n(x) \, Y$ is said %to be {\emph{coscalar, covector or cotensor }} of  power $\Pi(Y) =n$, this is called {\emph{scale covariance}}
%}
%

Summarizing, according to SIV, $a_0$ is not a universal constant, but depends on time, $H_0$, and $\Omega_m$.

\bibliographystyle{apsrev4-1}
\bibliography{saburova,Burkert}

\setlength\LTcapwidth{\textheight}
%\begin{landscape}
\scriptsize
%\tiny, \scriptsize, footnotesize,small, nrmalsize, large, Large, LARGE, huge, HUGE
	\begin{longtable*}{ccccccccccc}
	%{cc r@{ $\pm$ }l r@{ $\pm$ }l r@{ $\pm$ }l r@{ $\pm$ }l r@{ $\pm$ }l r@{ $\pm$ }l r@{ $\pm$ %}l r@{ $\pm$ }l r@{ $\pm$ }l c}
		\caption{\label{sample} 
(1) Galaxy name. (2) Best fit for $A_0 = \log_{10} a_0$ ($\rm km/s^2$). (3–8) $A_0$ credible intervals ($1 \sigma$, $3 \sigma$, $5 \sigma$). (9-10) Distance (Mpc), Error. (11) $Z=\log_{10} z $.
%
%		Results. The columns are: (1) Galaxy name. (2) Best fit for $A_0 = \log_{10} a_0$ ($\rm km/s^2$). (3–8) The limits of %the $A_0$ credible intervals, with respect to the A0 mode, for $1 \sigma$, $3 \sigma$, and $5 \sigma$. (9-10) Distance in %Mpc, and error. (11) $Z=\log_{10} z $.
%
}\\
		\hline	
    Galaxy      & \multicolumn{1}{l}{$A_0$ best } & \multicolumn{1}{l}{$1 \sigma_{-}$} & \multicolumn{1}{l}{$1 \sigma_{+}$  } & \multicolumn{1}{l}{$3 \sigma_{-}$   } & \multicolumn{1}{l}{$3 \sigma_{+}$  } & \multicolumn{1}{l}{$5 \sigma_{-}$   } & \multicolumn{1}{l}{$5 \sigma_{+}$  } & \multicolumn{1}{l}{D} & \multicolumn{1}{l}{$\rm D_{\rm Error}$} & \multicolumn{1}{l}{Z}\\
\hline  
\endfirsthead
\caption{Continued}\\
		\hline
		Galaxy      & \multicolumn{1}{l}{$A_0$ best } & \multicolumn{1}{l}{$1 \sigma_{-}$} & \multicolumn{1}{l}{$1 \sigma_{+}$  } & \multicolumn{1}{l}{$3 \sigma_{-}$   } & \multicolumn{1}{l}{$3 \sigma_{+}$  } & \multicolumn{1}{l}{$5 \sigma_{-}$   } & \multicolumn{1}{l}{$5 \sigma_{+}$  } & \multicolumn{1}{l}{D} & \multicolumn{1}{l}{$\rm D_{\rm Error}$} & \multicolumn{1}{l}{Z}\\
		\hline
	\endhead
    CamB        & -14.645 & -0.614 & 0.462 & -5.328 & 0.727 & -5.328 & 1.287 & 3.36  & 0.26  & -3.1237 \\
    D512-2      & -13.241 & -0.416 & 0.463 & -1.388 & 1.963 & -3.752 & 4.179 & 15.2  & 4.56  & -2.4682 \\
    D564-8      & -13.486 & -0.147 & 0.164 & -0.424 & 0.585 & -0.71 & 1.212 & 8.79  & 0.28  & -2.7061 \\
    D631-7      & -13.14 & -0.075 & 0.08  & -0.214 & 0.259 & -0.337 & 0.462 & 7.72  & 0.18  & -2.7625 \\
    DDO064      & -13.03 & -0.317 & 0.363 & -0.935 & 1.609 & -1.679 & 2.535 & 6.80  & 2.04  & -2.8176 \\
    DDO154      & -13.007 & -0.066 & 0.07  & -0.187 & 0.222 & -0.298 & 0.377 & 4.04  & 20\%  & -3.0437 \\
    DDO161      & -11.576 & -0.438 & 0.641 & -0.996 & 1.505 & -1.401 & 2.143 & 7.50  & 2.25  & -2.775 \\
    DDO168      & -12.993 & -0.146 & 0.168 & -0.388 & 0.592 & -0.577 & 1.157 & 4.25  & 0.21  & -3.0217 \\
    DDO170      & -13.615 & -0.245 & 0.288 & -0.657 & 1.086 & -0.985 & 1.794 & 15.40 & 4.62  & -2.4626 \\
    ESO079-G014 & -12.785 & -0.242 & 0.284 & -0.654 & 1.137 & -1.008 & 1.776 & 28.70 & 7.17  & -2.1922 \\
    ESO116-G012 & -12.479 & -0.273 & 0.362 & -0.651 & 1.706 & -0.945 & 2.37  & 13.00 & 3.90  & -2.5361 \\
    ESO444-G084 & -12.322 & -0.308 & 0.37  & -0.807 & 1.457 & -1.153 & 2.589 & 4.83  & 0.48  & -2.9661 \\
    ESO563-G021 & -12.866 & -0.088 & 0.094 & -0.253 & 0.301 & -0.404 & 0.535 & 60.8  & 9.10  & -1.8662 \\
    F565-V2     & -12.742 & -0.258 & 0.306 & -0.687 & 1.149 & -1.011 & 2.185 & 51.8  & 10.00 & -1.9357 \\
    F568-3      & -13.065 & -0.381 & 0.468 & -1.09 & 2.097 & -2.011 & 3.371 & 82.40 & 8.24  & -1.7341 \\
    F568-V1     & -12.544 & -0.352 & 0.441 & -1.056 & 2.656 & -1.503 & 4.102 & 80.60 & 8.06  & -1.7437 \\
    F571-8      & -11.482 & -0.22 & 0.281 & -0.557 & 1.596 & -0.783 & 2.307 & 53.30 & 10.70 & -1.9233 \\
    F571-V1     & -13.074 & -0.531 & 0.674 & -1.317 & 4.444 & -6.797 & 6.756 & 80.10 & 8.00  & -1.7464 \\
    F574-1      & -13.03 & -0.196 & 0.22  & -0.572 & 0.841 & -0.975 & 1.564 & 96.80 & 9.68  & -1.6642 \\
    F583-1      & -12.998 & -0.254 & 0.293 & -0.763 & 1.384 & -1.067 & 2.574 & 35.40 & 8.85  & -2.1011 \\
    F583-4      & -13.091 & -0.321 & 0.364 & -0.937 & 1.353 & -1.806 & 2.401 & 53.30 & 10.70 & -1.9233 \\
    IC2574      & -13.099 & -0.067 & 0.08  & -0.177 & 0.32  & -0.26 & 0.56  & 3.91  & 0.20  & -3.0579 \\
    IC4202      & -12.607 & -0.452 & 2.562 & -0.612 & 4.467 & -1.012 & 4.686 & 100.40 & 10.00 & -1.6483 \\
    KK98-251    & -13.834 & -0.376 & 0.395 & -5.703 & 1.539 & -2.491 & 3.983 & 6.80  & 2.04  & -2.8176 \\
    NGC0024     & -12.862 & -0.095 & 0.097 & -0.277 & 0.297 & -0.459 & 0.515 & 7.30  & 0.36  & -2.7868 \\
    NGC0055     & -13.205 & -0.067 & 0.069 & -0.201 & 0.212 & -0.332 & 0.372 & 2.11  & 0.11  & -3.3258 \\
    NGC0100     & -12.603 & -0.282 & 0.368 & -0.702 & 1.706 & -1.027 & 2.507 & 13.50 & 4.05  & -2.5197 \\
    NGC0247     & -13.451 & -0.135 & 0.126 & -0.448 & 0.356 & -0.815 & 0.56  & 3.70  & 0.19  & -3.0819 \\
    NGC0289     & -13.489 & -0.165 & 0.17  & -0.49 & 0.535 & -0.773 & 0.891 & 20.80 & 5.20  & -2.332 \\
    NGC0300     & -12.493 & -0.351 & 0.47  & -0.8  & 2.343 & -1.057 & 3.43  & 2.08  & 0.10  & -3.332 \\
    NGC0801     & -13.546 & -0.059 & 0.061 & -0.175 & 0.191 & -0.283 & 0.342 & 80.70 & 8.07  & -1.7432 \\
    NGC0891     & -12.612 & -0.051 & 0.052 & -0.148 & 0.16  & -0.243 & 0.281 & 9.91  & 0.50  & -2.654 \\
    NGC1003     & -10.827 & -0.516 & 0.661 & -1.291 & 1.168 & -1.726 & 1.47  & 11.40 & 3.42  & -2.5932 \\
    NGC1090     & -13.181 & -0.14 & 0.151 & -0.4  & 0.491 & -0.629 & 0.784 & 37.00 & 9.25  & -2.0819 \\
    NGC2403     & -12.006 & -0.09 & 0.095 & -0.256 & 0.305 & -0.368 & 0.462 & 3.16  & 0.16  & -3.1504 \\
    NGC2683     & -13.241 & -0.096 & 0.096 & -0.287 & 0.295 & -0.489 & 0.496 & 9.81  & 0.49  & -2.6584 \\
    NGC2841     & -12.898 & -0.055 & 0.058 & -0.156 & 0.188 & -0.245 & 0.324 & 14.10 & 1.40  & -2.5009 \\
    NGC2903     & -12.317 & -0.104 & 0.106 & -0.304 & 0.317 & -0.437 & 0.462 & 6.60  & 1.98  & -2.8305 \\
    NGC2915     & -12.399 & -0.104 & 0.113 & -0.296 & 0.369 & -0.475 & 0.703 & 4.06  & 0.20  & -3.0415 \\
    NGC2955     & -12.903 & -0.134 & 0.139 & -0.39 & 0.433 & -0.633 & 0.727 & 97.9  & 9.80  & -1.6593 \\
    NGC2976     & -18.324 & -2.469 & 1.421 & -3.109 & 3.518 & -3.109 & 4.828 & 3.58  & 0.18  & -3.0962 \\
    NGC2998     & -13.369 & -0.1  & 0.105 & -0.29 & 0.338 & -0.471 & 0.613 & 68.1  & 10.2  & -1.8169 \\
    NGC3109     & -12.76 & -0.075 & 0.084 & -0.206 & 0.285 & -0.318 & 0.529 & 1.33  & 0.07  & -3.5262 \\
    NGC3198     & -12.92 & -0.094 & 0.101 & -0.264 & 0.334 & -0.392 & 0.511 & 13.8  & 1.40  & -2.5102 \\
    NGC3521     & -12.928 & -0.115 & 0.119 & -0.332 & 0.371 & -0.529 & 0.604 & 7.70  & 2.30  & -2.7636 \\
    NGC3726     & -13.133 & -0.149 & 0.159 & -0.428 & 0.526 & -0.682 & 0.984 & 18.00 & 2.50  & -2.3948 \\
    NGC3741     & -12.828 & -0.07 & 0.076 & -0.2  & 0.246 & -0.31 & 0.441 & 3.21  & 0.17  & -3.1436 \\
    NGC3769     & -12.98 & -0.141 & 0.153 & -0.402 & 0.519 & -0.626 & 1.052 & 18.00 & 2.50  & -2.3948 \\
    NGC3877     & -13.184 & -0.745 & 0.577 & -6.789 & 0.834 & -6.789 & 1.545 & 18.00 & 2.50  & -2.3948 \\
    NGC3893     & -12.737 & -0.139 & 0.148 & -0.402 & 0.486 & -0.639 & 0.908 & 18.00 & 2.50  & -2.3948 \\
    NGC3917     & -13.232 & -0.184 & 0.186 & -0.559 & 0.581 & -0.973 & 1.076 & 18.00 & 2.50  & -2.3948 \\
    NGC3949     & -12.935 & -0.32 & 0.292 & -6.868 & 0.889 & -7.053 & 1.382 & 18.00 & 2.50  & -2.3948 \\
    NGC3953     & -13.77 & -3.868 & 0.454 & -6.15 & 0.806 & -6.15 & 1.581 & 18.00 & 2.50  & -2.3948 \\
    NGC3972     & -12.766 & -0.19 & 0.19  & -0.6  & 0.608 & -1.073 & 1.077 & 18.00 & 2.50  & -2.3948 \\
    NGC3992     & -13.299 & -0.099 & 0.102 & -0.29 & 0.318 & -0.479 & 0.568 & 23.70 & 2.30  & -2.2753 \\
    NGC4010     & -12.84 & -0.168 & 0.175 & -0.501 & 0.58  & -0.839 & 1.128 & 18.00 & 2.50  & -2.3948 \\
    NGC4013     & -12.858 & -0.083 & 0.089 & -0.233 & 0.29  & -0.365 & 0.521 & 18.00 & 2.50  & -2.3948 \\
    NGC4051     & -13.571 & -5.087 & 0.431 & -6.366 & 0.933 & -6.366 & 1.766 & 18.00 & 2.50  & -2.3948 \\
    NGC4068     & -13.607 & -0.335 & 0.331 & -5.465 & 1.17  & -6.357 & 1.93  & 4.37  & 0.22  & -3.0096 \\
    NGC4085     & -12.829 & -0.212 & 0.212 & -0.682 & 0.682 & -1.417 & 1.396 & 18.00 & 2.50  & -2.3948 \\
    NGC4088     & -13.149 & -0.147 & 0.155 & -0.426 & 0.512 & -0.707 & 0.953 & 18.00 & 2.50  & -2.3948 \\
    NGC4100     & -13.182 & -0.094 & 0.099 & -0.273 & 0.315 & -0.445 & 0.573 & 18.00 & 2.50  & -2.3948 \\
    NGC4138     & -13.069 & -0.213 & 0.212 & -0.686 & 0.657 & -1.623 & 1.21  & 18.00 & 2.50  & -2.3948 \\
    NGC4157     & -12.978 & -0.121 & 0.13  & -0.343 & 0.433 & -0.543 & 0.779 & 18.00 & 2.50  & -2.3948 \\
    NGC4183     & -13.506 & -0.145 & 0.151 & -0.425 & 0.485 & -0.704 & 0.89  & 18.00 & 2.50  & -2.3948 \\
    NGC4217     & -12.419 & -0.181 & 0.208 & -0.496 & 0.815 & -0.745 & 1.504 & 18.00 & 2.50  & -2.3948 \\
    NGC4559     & -12.954 & -0.258 & 0.318 & -0.689 & 1.42  & -0.96 & 2.025 & 9.00  & 2.70  & -2.6958 \\
    NGC5005     & -12.783 & -0.28 & 0.246 & -6.503 & 0.704 & -7.205 & 1.001 & 16.90 & 1.50  & -2.4222 \\
    NGC5033     & -12.933 & -0.1  & 0.101 & -0.295 & 0.307 & -0.442 & 0.47  & 15.70 & 4.70  & -2.4542 \\
    NGC5055     & -12.821 & -0.098 & 0.098 & -0.29 & 0.293 & -0.426 & 0.444 & 9.90  & 0.30  & -2.6544 \\
    NGC5371     & -13.769 & -0.09 & 0.092 & -0.265 & 0.283 & -0.447 & 0.437 & 39.70 & 9.92  & -2.0513 \\
    NGC5585     & -12.21 & -0.37 & 0.552 & -0.764 & 1.895 & -1.156 & 2.194 & 7.06  & 2.12  & -2.8013 \\
    NGC5907     & -13.373 & -0.04 & 0.041 & -0.117 & 0.125 & -0.186 & 0.211 & 17.30 & 0.90  & -2.412 \\
    NGC5985     & -13.567 & -0.095 & 0.098 & -0.28 & 0.31  & -0.467 & 0.556 & 39.70 & 9.90  & -2.0513 \\
    NGC6015     & -13.536 & -0.094 & 0.098 & -0.272 & 0.306 & -0.443 & 0.498 & 17.00 & 5.10  & -2.4196 \\
    NGC6195     & -13.133 & -0.105 & 0.108 & -0.312 & 0.329 & -0.501 & 0.562 & 127.80 & 12.80 & -1.5435 \\
    NGC6503     & -12.827 & -0.043 & 0.045 & -0.126 & 0.142 & -0.201 & 0.239 & 6.26  & 0.31  & -2.8535 \\
    NGC6674     & -13.504 & -0.082 & 0.085 & -0.238 & 0.264 & -0.39 & 0.452 & 51.20 & 10.20 & -1.9408 \\
    NGC6789     & -12.214 & -0.284 & 0.314 & -0.841 & 1.141 & -1.492 & 2.393 & 3.52  & 0.18  & -3.1035 \\
    NGC6946     & -13.033 & -0.098 & 0.099 & -0.293 & 0.3   & -0.466 & 0.462 & 5.52  & 1.66  & -2.9081 \\
    NGC7331     & -12.785 & -0.063 & 0.066 & -0.18 & 0.211 & -0.275 & 0.353 & 14.70 & 1.50  & -2.4828 \\
    NGC7793     & -13.243 & -0.205 & 0.23  & -0.57 & 0.823 & -0.926 & 1.676 & 3.61  & 0.18  & -3.0926 \\
    NGC7814     & -12.572 & -0.058 & 0.058 & -0.18 & 0.171 & -0.307 & 0.284 & 14.40 & 0.66  & -2.4917 \\
    UGC00128    & -13.691 & -0.116 & 0.128 & -0.32 & 0.429 & -0.487 & 0.735 & 64.50 & 9.70  & -1.8405 \\
    UGC00191    & -13.65 & -0.241 & 0.259 & -0.707 & 0.887 & -1.159 & 1.378 & 17.10 & 5.10  & -2.4171 \\
    UGC00634    & -12.969 & -0.379 & 0.45  & -1.014 & 1.701 & -1.469 & 2.544 & 30.90 & 7.70  & -2.1601 \\
    UGC00731    & -13.257 & -0.32 & 0.369 & -0.98 & 1.834 & -2.166 & 2.757 & 12.50 & 3.75  & -2.5532 \\
    UGC00891    & -12.334 & -0.379 & 0.554 & -0.8  & 2.19  & -1.214 & 2.594 & 10.20 & 3.10  & -2.6415 \\
    UGC01281    & -12.981 & -0.083 & 0.08  & -0.263 & 0.233 & -0.476 & 0.384 & 5.27  & 0.24  & -2.9283 \\
    UGC02259    & -13.736 & -0.236 & 0.251 & -0.705 & 0.85  & -1.283 & 1.635 & 10.50 & 3.10  & -2.6289 \\
    UGC02487    & -13.349 & -0.074 & 0.075 & -0.22 & 0.229 & -0.357 & 0.377 & 69.10 & 10.40 & -1.8106 \\
    UGC02885    & -12.964 & -0.101 & 0.105 & -0.292 & 0.33  & -0.463 & 0.577 & 80.60 & 8.06  & -1.7437 \\
    UGC02916    & -13.551 & -0.156 & 0.153 & -0.515 & 0.465 & -1.21 & 0.758 & 65.40 & 9.80  & -1.8345 \\
    UGC02953    & -13.032 & -0.067 & 0.069 & -0.2  & 0.211 & -0.281 & 0.338 & 16.50 & 4.95  & -2.4326 \\
    UGC03205    & -13.201 & -0.085 & 0.088 & -0.245 & 0.275 & -0.39 & 0.461 & 50.00 & 10.00 & -1.9511 \\
    UGC03546    & -12.884 & -0.103 & 0.107 & -0.313 & 0.352 & -0.453 & 0.639 & 28.70 & 7.20  & -2.1922 \\
    UGC03580    & -10.529 & -0.525 & 0.646 & -1.473 & 1.102 & -2.141 & 1.434 & 20.70 & 5.20  & -2.3341 \\
    UGC04278    & -12.504 & -0.319 & 0.426 & -0.766 & 2.138 & -1.135 & 2.641 & 9.51  & 2.85  & -2.6719 \\
    UGC04325    & -13.656 & -5.926 & 0.625 & -6.163 & 1.357 & -6.163 & 3.269 & 9.60  & 2.88  & -2.6678 \\
    UGC04483    & -13.395 & -0.141 & 0.142 & -0.428 & 0.434 & -0.733 & 0.759 & 3.34  & 0.31  & -3.1263 \\
    UGC04499    & -13.231 & -0.297 & 0.344 & -0.872 & 1.609 & -1.371 & 3.036 & 12.50 & 3.75  & -2.5532 \\
    UGC05005    & -13.197 & -0.419 & 0.501 & -1.185 & 2.231 & -1.859 & 4.122 & 53.70 & 10.70 & -1.9201 \\
    UGC05253    & -12.707 & -0.085 & 0.086 & -0.255 & 0.259 & -0.398 & 0.411 & 22.90 & 5.72  & -2.2902 \\
    UGC05414    & -13.154 & -0.322 & 0.371 & -1.043 & 1.974 & -2.442 & 2.8   & 9.40  & 2.82  & -2.6769 \\
    UGC05716    & -13.273 & -0.285 & 0.343 & -0.732 & 1.225 & -1.11 & 1.652 & 21.30 & 5.30  & -2.3217 \\
    UGC05721    & -12.573 & -0.241 & 0.291 & -0.641 & 1.266 & -0.906 & 2.059 & 6.18  & 1.85  & -2.8591 \\
    UGC05750    & -13.355 & -0.285 & 0.313 & -0.861 & 1.124 & -2.09 & 2.168 & 58.70 & 11.70 & -1.8814 \\
    UGC05764    & -14.603 & -1.104 & 2.539 & -1.545 & 4.629 & -1.545 & 5.561 & 7.47  & 2.24  & -2.7768 \\
    UGC05829    & -13.033 & -0.602 & 0.717 & -2.34 & 4.449 & -6.898 & 5.976 & 8.64  & 2.59  & -2.7136 \\
    UGC05918    & -13.083 & -0.334 & 0.38  & -0.979 & 1.702 & -1.592 & 2.999 & 7.66  & 2.30  & -2.7658 \\
    UGC05986    & -12.436 & -0.24 & 0.306 & -0.605 & 1.531 & -0.851 & 2.004 & 8.63  & 2.59  & -2.7141 \\
    UGC06399    & -12.847 & -0.167 & 0.172 & -0.51 & 0.562 & -0.906 & 1.109 & 18.00 & 2.50  & -2.3948 \\
    UGC06446    & -13.18 & -0.266 & 0.301 & -0.76 & 1.177 & -1.187 & 1.956 & 12.00 & 3.60  & -2.5709 \\
    UGC06614    & -12.415 & -0.372 & 0.479 & -0.968 & 2.554 & -1.336 & 4.065 & 88.70 & 8.87  & -1.7022 \\
    UGC06667    & -12.561 & -0.136 & 0.149 & -0.388 & 0.516 & -0.608 & 1.042 & 18.00 & 2.50  & -2.3948 \\
    UGC06786    & -12.087 & -0.158 & 0.178 & -0.418 & 0.606 & -0.597 & 0.909 & 29.30 & 7.32  & -2.1832 \\
    UGC06787    & -13.387 & -0.048 & 0.05  & -0.139 & 0.156 & -0.218 & 0.277 & 21.30 & 5.32  & -2.3217 \\
    UGC06818    & -13.089 & -0.18 & 0.184 & -0.549 & 0.595 & -0.935 & 1.175 & 18.00 & 2.50  & -2.3948 \\
    UGC06917    & -12.882 & -0.166 & 0.172 & -0.491 & 0.567 & -0.819 & 1.112 & 18.00 & 2.50  & -2.3948 \\
    UGC06923    & -13.031 & -0.196 & 0.195 & -0.63 & 0.622 & -1.401 & 1.232 & 18.00 & 2.50  & -2.3948 \\
    UGC06930    & -13.382 & -0.356 & 0.37  & -1.354 & 1.399 & -4.689 & 3.052 & 18.00 & 2.50  & -2.3948 \\
    UGC06983    & -13.01 & -0.149 & 0.158 & -0.432 & 0.524 & -0.7  & 1.021 & 18.00 & 2.50  & -2.3948 \\
    UGC07089    & -13.308 & -0.189 & 0.189 & -0.604 & 0.609 & -1.175 & 1.143 & 18.00 & 2.50  & -2.3948 \\
    UGC07125    & -13.871 & -0.285 & 0.327 & -0.877 & 1.551 & -1.516 & 2.282 & 19.80 & 5.90  & -2.3534 \\
    UGC07151    & -13.265 & -0.131 & 0.12  & -0.437 & 0.337 & -0.837 & 0.524 & 6.87  & 0.34  & -2.8131 \\
    UGC07232    & -12.847 & -0.213 & 0.203 & -0.726 & 0.611 & -3.71 & 1.102 & 2.83  & 0.17  & -3.1983 \\
    UGC07261    & -13.236 & -0.719 & 0.839 & -6.698 & 3.382 & -6.698 & 6.573 & 13.10 & 3.93  & -2.5328 \\
    UGC07323    & -13.118 & -0.383 & 0.407 & -2.974 & 2.245 & -6.843 & 2.622 & 8.00  & 2.40  & -2.747 \\
    UGC07399    & -12.272 & -0.252 & 0.308 & -0.676 & 1.453 & -0.957 & 2.208 & 8.43  & 2.53  & -2.7242 \\
    UGC07524    & -13.326 & -0.142 & 0.143 & -0.436 & 0.438 & -0.765 & 0.751 & 4.74  & 0.24  & -2.9743 \\
    UGC07559    & -13.614 & -0.167 & 0.155 & -0.572 & 0.444 & -1.645 & 0.79  & 4.97  & 0.25  & -2.9537 \\
    UGC07577    & -14.597 & -3.045 & 0.474 & -5.333 & 0.813 & -5.333 & 1.459 & 2.59  & 0.13  & -3.2368 \\
    UGC07603    & -12.501 & -0.266 & 0.346 & -0.659 & 1.715 & -0.943 & 2.149 & 4.70  & 1.41  & -2.978 \\
    UGC07690    & -13.541 & -0.389 & 0.404 & -6.31 & 1.609 & -6.454 & 2.388 & 8.11  & 2.43  & -2.7411 \\
    UGC07866    & -13.476 & -0.269 & 0.265 & -0.935 & 0.861 & -6.501 & 1.448 & 4.57  & 0.23  & -2.9902 \\
    UGC08286    & -12.989 & -0.061 & 0.06  & -0.189 & 0.176 & -0.323 & 0.29  & 6.50  & 0.21  & -2.8372 \\
    UGC08490    & -12.963 & -0.119 & 0.128 & -0.34 & 0.413 & -0.54 & 0.735 & 4.65  & 0.53  & -2.9826 \\
    UGC08550    & -12.975 & -0.241 & 0.302 & -0.622 & 1.329 & -0.885 & 2.039 & 6.70  & 2.00  & -2.824 \\
    UGC08699    & -12.978 & -0.122 & 0.137 & -0.34 & 0.479 & -0.521 & 0.861 & 39.30 & 9.82  & -2.0557 \\
    UGC08837    & -13.599 & -0.129 & 0.12  & -0.439 & 0.35  & -0.909 & 0.588 & 7.21  & 0.36  & -2.7921 \\
    UGC09037    & -12.876 & -0.136 & 0.146 & -0.382 & 0.478 & -0.588 & 0.839 & 83.6  & 8.40  & -1.7279 \\
    UGC09133    & -13.146 & -0.065 & 0.065 & -0.192 & 0.197 & -0.318 & 0.297 & 57.10 & 11.40 & -1.8934 \\
    UGC09992    & -13.996 & -5.379 & 0.911 & -5.672 & 3.04  & -5.672 & 7.389 & 10.70 & 3.21  & -2.6207 \\
    UGC10310    & -13.29 & -0.772 & 0.746 & -6.666 & 1.808 & -6.666 & 4.33  & 15.20 & 4.60  & -2.4682 \\
    UGC11455    & -12.841 & -0.095 & 0.103 & -0.268 & 0.337 & -0.416 & 0.58  & 78.60 & 11.80 & -1.7547 \\
    UGC11557    & -13.209 & -1.105 & 1.227 & -6.676 & 4.01  & -6.676 & 6.456 & 24.20 & 6.05  & -2.2663 \\
    UGC11820    & -13.069 & -0.379 & 0.47  & -0.938 & 1.757 & -1.34 & 2.426 & 18.10 & 5.43  & -2.3924 \\
    UGC11914    & -12.274 & -0.128 & 0.129 & -0.38 & 0.384 & -0.61 & 0.575 & 16.90 & 5.10  & -2.4222 \\
    UGC12506    & -13.546 & -0.116 & 0.117 & -0.355 & 0.359 & -0.624 & 0.625 & 100.60 & 10.10 & -1.6475 \\
    UGC12632    & -13.674 & -0.35 & 0.361 & -2.922 & 1.682 & -6.297 & 2.197 & 9.77  & 2.93  & -2.6602 \\
    UGC12732    & -13.386 & -0.356 & 0.399 & -1.003 & 1.481 & -1.743 & 2.227 & 13.20 & 4.00  & -2.5295 \\
    UGCA442     & -12.786 & -0.133 & 0.165 & -0.329 & 0.629 & -0.462 & 1.305 & 4.35  & 0.22  & -3.0116 \\
    UGCA444     & -13.26 & -0.077 & 0.078 & -0.231 & 0.242 & -0.381 & 0.415 & 0.98  & 0.05  & -3.6588 \\    
\hline    
    \end{longtable*}
%\end{landscape}

\end{document}